\documentclass[preprint]{aastex}

\usepackage{color}
\usepackage{multicol}
\usepackage{multirow}
\usepackage{subfigure}
\usepackage{float}
\usepackage{hyperref}
\usepackage[percent]{overpic}
\usepackage{threeparttable}

\newcommand{\rhessi}[0]{{\it RHESSI}}
\newcommand{\goes}[0]{{\it GOES}}
\newcommand{\nustar}[0]{{\it NuSTAR}}
\newcommand{\foxsi}[0]{{\it FOXSI}}

\newcommand{\app}[0]{$\sim$}

\begin{document}

\title{Accelerated electrons observed down to $<$7 keV in a \nustar\ solar microflare}

\author{Lindsay Glesener\altaffilmark{1}, S\"{a}m Krucker\altaffilmark{2,3}, Jessie Duncan\altaffilmark{1}, Iain G. Hannah\altaffilmark{4}, Brian W. Grefenstette\altaffilmark{5}, Bin Chen\altaffilmark{6}, David M. Smith\altaffilmark{7}, Stephen M. White\altaffilmark{8}, Hugh Hudson\altaffilmark{2,4}
 }

\altaffiltext{1}{University of Minnesota, Minneapolis, USA}
\altaffiltext{2}{University of California at Berkeley, Berkeley, USA}
\altaffiltext{3}{University of Applied Sciences and Arts Northwestern Switzerland, Windisch, Switzerland}
\altaffiltext{4}{University of Glasgow, Glasgow, UK}
\altaffiltext{5}{California Institute of Technology, Pasadena, USA}
\altaffiltext{6}{New Jersey Institute of Technology, Newark, USA}
\altaffiltext{7}{University of California at Santa Cruz, Santa Cruz, USA}
\altaffiltext{8}{Air Force Research Laboratory, Albuquerque, USA}

\begin{abstract}

We report the detection of emission from a non-thermal electron distribution in a small solar microflare (\goes\ class A5.7) observed by the \textit{Nuclear Spectroscopic Telescope Array (NuSTAR)}, with supporting observation by the \textit{Reuven Ramaty High Energy Solar Spectroscopic Imager (RHESSI)}. 
  The flaring plasma is well accounted for by a thick-target model of accelerated electrons collisionally thermalizing within the loop, akin to the ``coronal thick target'' behavior occasionally observed in larger flares.  This is the first positive detection of non-thermal hard X-rays from the Sun using a direct imager (as opposed to indirectly imaging instruments).  The accelerated electron distribution has a spectral index of 6.3$\pm$0.7, extends down to at least 6.5 keV, and deposits energy at a rate of $\sim 2\times$10$^{27}$ erg s$^{-1}$, heating the flare loop to at least 10 MK.  The existence of dominant non-thermal emission in X-rays down to $<$5 keV means that \rhessi\ emission is almost entirely non-thermal, contrary to what is usually assumed in \rhessi\ spectroscopy.  The ratio of non-thermal to thermal energies is similar to that of large flares, in contrast to what has been found in previous studies of small \rhessi\ flares.  We suggest that a coronal thick target may be a common property of many small microflares based on the average electron energy and collisional mean free path.  Future observations of this kind will enable understanding of how flare particle acceleration changes across energy scales, and will aid the push toward the observational regime of nanoflares, which are a possible source of significant coronal heating.

\end{abstract}

\section{Introduction}

While large solar flares of classes M and X garner most public and scientific attention, small flares occur far more frequently.  Microflares of \goes\  class A and B occur exclusively in active regions \citep[e.g.][]{hannah2008, christe2008}.  Small, sub-A-class brightenings have been observed in active and in quiet regions \citep[e.g.][]{kuhar2018}, but it is still unclear as to whether faint quiet-Sun brightenings signify the same physical processes as those that occur in active regions.  Microflares are generally observed to be similar in nature to larger flares, with impulsive phases followed by gradual cooling \citep[e.g.][]{glesener2017}.

There is a great deal of interest in studying the parameters of flare-accelerated electrons across a wide range of flare magnitudes, including to very small flare energies.  The observed scaling properties can help to assess theories of particle acceleration.  In particular, since nanoflares are possible coronal heating candidates, it is of great interest to examine particle acceleration in small events to determine if they are similar to large flares and how much energy they could deposit in the corona.  One factor that could affect the efficiency of particle acceleration across energy scales is the magnitude of the guide field relative to the reconnecting field.  \citet{dahlin2016} and \citet{dahlin2017} have shown that the presence of a large-order guide field (greater than the reconnecting field) suppresses particle acceleration, which would lead one to expect low acceleration efficiency for nanoflares.  While the small flares currently observable in hard X-rays (HXRs) are still far from the nanoflare scale, each advance in sensitivity pushes further toward this observational regime.

HXR observations are ideal for characterizing the hot thermal plasma and any non-thermal emissions generated in small energy releases.  However, the most advanced solar HXR spacecraft instrument, the \textit{Reuven Ramaty High Energy Solar Spectroscopic Imager (RHESSI)} spacecraft, was limited in its sensitivity to small flares due to its indirect imaging method, although \goes\  class A microflares could be observed \citep{hannah2008, christe2008}.  
  The recent advent of directly focusing HXR instruments in the form of the \textit{Nuclear Spectroscopic Telescope Array (NuSTAR)} spacecraft and the \textit{Focusing Optics X-ray Solar Imager (FOXSI)} sounding rocket have enabled the observation of small microflares orders of magnitude fainter than those observed by \rhessi\  \citep{krucker2014, glesener2016, glesener2017, wright2017, kuhar2018, athiray2020, hannah2019, vievering2019}.
    However, while some of these observations implied a high-energy excess that might arise from accelerated electron distributions, a clear, distinct measurement of flare-accelerated electrons by focusing HXR instruments has heretofore been prevented by the faintness of these flares and the rarity of observational opportunity (since neither \nustar\ nor \foxsi\ observes the Sun often).

Here we report the first direct detection of non-thermal emission from a solar flare using a focusing HXR imager.  We analyze the microflare's thermal and non-thermal properties and compare these to larger flares.

\section{Observations}

\nustar\, is a NASA Astrophysics Small Explorer launched in 2012 \citep{harrison2013}.  Unlike all previous HXR-observing spacecraft, \nustar\ utilizes directly focusing HXR optics to achieve far better sensitivity than any previous HXR instrument.  Although it is an astrophysics mission, \nustar\ can measure faint solar emission during relatively quiet times, when best use is made of the instrument's limited throughput \citep{grefenstette2016}.  Observations of the Sun are performed several times per year for one to several hours at a time.

On 2017 August 21, \nustar\ observed the Sun for one orbit just before the solar disk was partly occulted by the Moon (\nustar\ observation IDs 20312001001 and 20312002001).  For a few minutes on this day, the Sun was totally occulted as viewed from several locations in North America, an event commonly referred to as the ``Great American Eclipse.''  Due to the high level of public excitement and scientific interest generated by this event, most telescopes capable of observing the Sun did so on that day.  

\nustar's target of interest on 2017 August 21 was an active region with NOAA number 12671, which was observed from 18:49:58 to 19:50:03 UTC,\footnote{Summary plots and information can be found at \url{http://ianan.github.io/nsigh_all/}.} a total of 3605 s, ($\sim$2940 s before the lunar occultation of the region began at \app19:39).  \nustar's livetime during the non-occulted observation was, on average, 0.3\% (with a minimum of 0.1\% at microflare peak) for an effective exposure of \app9 s.  This region produced a few C-class flares before and after the \nustar\ observation and produced several evident microflares during the observation.  Here, we concentrate on a microflare occurring in the west of the active region, at a location approximately [360,45] arcsec west and north of the solar center.  (Future work will analyze in detail the entire set of \nustar\ microflares in this region.)

\subsection{Microflare temporal and spatial observations}

\begin{figure}[htbp]
\begin{center}
	\includegraphics[width=0.5\textwidth]{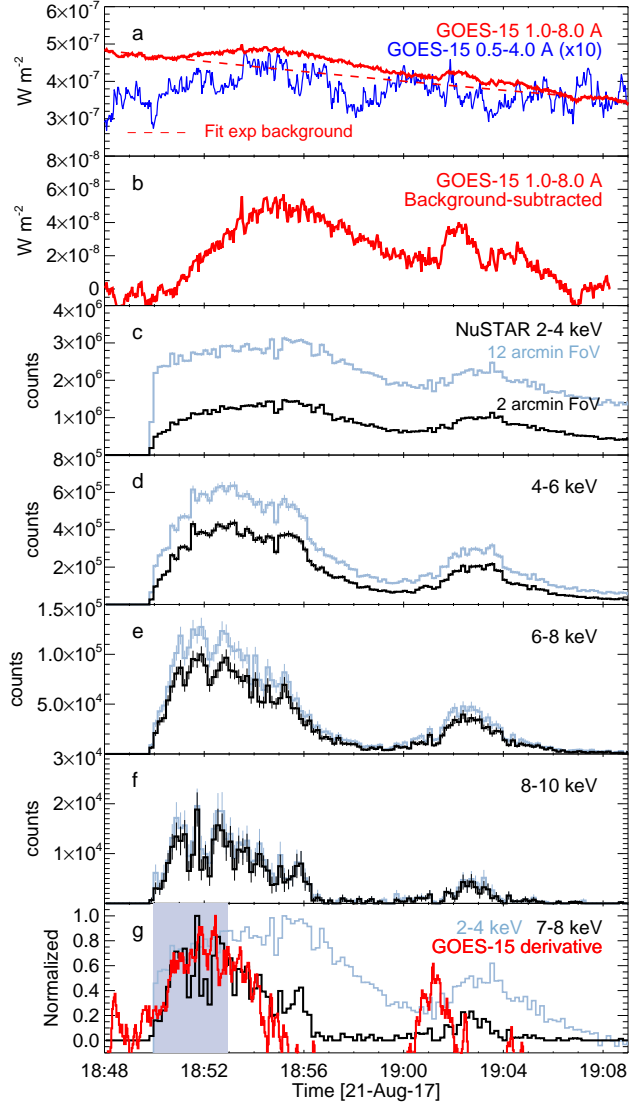}
\caption{ Lightcurves from \nustar\ and \goes.  Panels (a) and (b) show \goes\  soft X-ray flux, with an exponentially falling background subtracted for Panel (b).  In Panel (a), the shorter-wavelength \goes\ channel flux has been multiplied by a factor of 10.  Panels (c)-(f) show livetime-corrected \nustar\ counts from both telescopes in 10 s time bins and several energy bands, with 1-sigma statistical uncertainties shown.  Pale blue lines show data integrated over the entire active region, while black lines show a 2 arcmin FoV around the microflare site.    Panel (g) shows that the high-energy \nustar\ emission mimics the derivative of the \goes\  long wavelength flux in the first minutes of the flare, indicative of the Neupert effect.  The gray box indicates the 3 min interval chosen for spectroscopy, and the 2-4 keV background-subtracted curve is shown for comparison.
}
\label{fig:lightcurves}
\end{center}
\end{figure}

\nustar\ has two quasi-identical telescopes.  These record X-rays on a single-photon basis; events can then be arbitrarily binned in space, time, and energy.  Since \nustar\ pointing knowledge carries large uncertainties during solar observation \citep[see][]{grefenstette2016}, we coaligned \nustar\ images empirically to data from the Atmospheric Imaging Assembly (AIA) aboard the \textit{Solar Dynamics Observatory (SDO)}.  A linear combination of AIA data in the 94\AA, 171\AA, and 211\AA\ filters was taken to isolate the Fe XVIII contribution, as in \citet{delzanna2013}; this line has a formation temperature of log T $\approx$ 6.9 and is sensitive to a temperature range that overlaps that of \nustar.  We coaligned \nustar\ and AIA Fe XVIII data at the peak of the microflare (\app18:55 UTC) and then cross-correlated the \nustar\ images to each other; this method assumes slow (or no) source motion.  Uncertainties in this coalignment are estimated to be \app10 arcsec.

Figure \ref{fig:lightcurves} shows (Panels (a)-(b)) soft X-ray lightcurves from the \goes\ X-ray Spectrometer (XRS).  After subtracting an exponentially falling background from the long-wavelength flux, the \goes\ class is A5.7.  Panels (c)-(f) show \nustar\ lightcurves in several energy bands for a 2 arcmin region centered on the flare, and also for a 12 arcmin region that corresponds to \nustar's entire field of view (FoV).  The active region was contained well within this 12 arcmin FoV.  At lower \nustar\ energies, emission from the entire active region is evident, while at higher energies, emission emanates from the flare only.  (Residual differences between blue and black curves in the 8-10 keV energy band are due to the wings of the instrumental point spread function, half power diameter \app1 arcmin.)  High time variability is evident, especially at higher energies.  \nustar\ high-energy emission closely follows the derivative of the flux in the \goes\ low-energy channel in the first few minutes of the microflare (see Panel (g)).  In this panel, \goes\ data have been smoothed over a boxcar interval of 2 min before taking the derivative, and both \nustar\ and the \goes\ derivative have been normalized to their maximum values over the plot time range.  The gray box in Panel (g) shows a 3 min interval at the beginning of the flare (18:50--18:53 UT) on which we concentrate our efforts in this paper; this interval was chosen because it covers the impulsive phase of the flare and because the \nustar\ pointing was relatively steady over this interval; the last \app minute of the impulsive phase was excluded due to pointing motion, which would complicate analysis.

\begin{figure*}[htbp]
\begin{center}
	\includegraphics[width=0.6\linewidth]{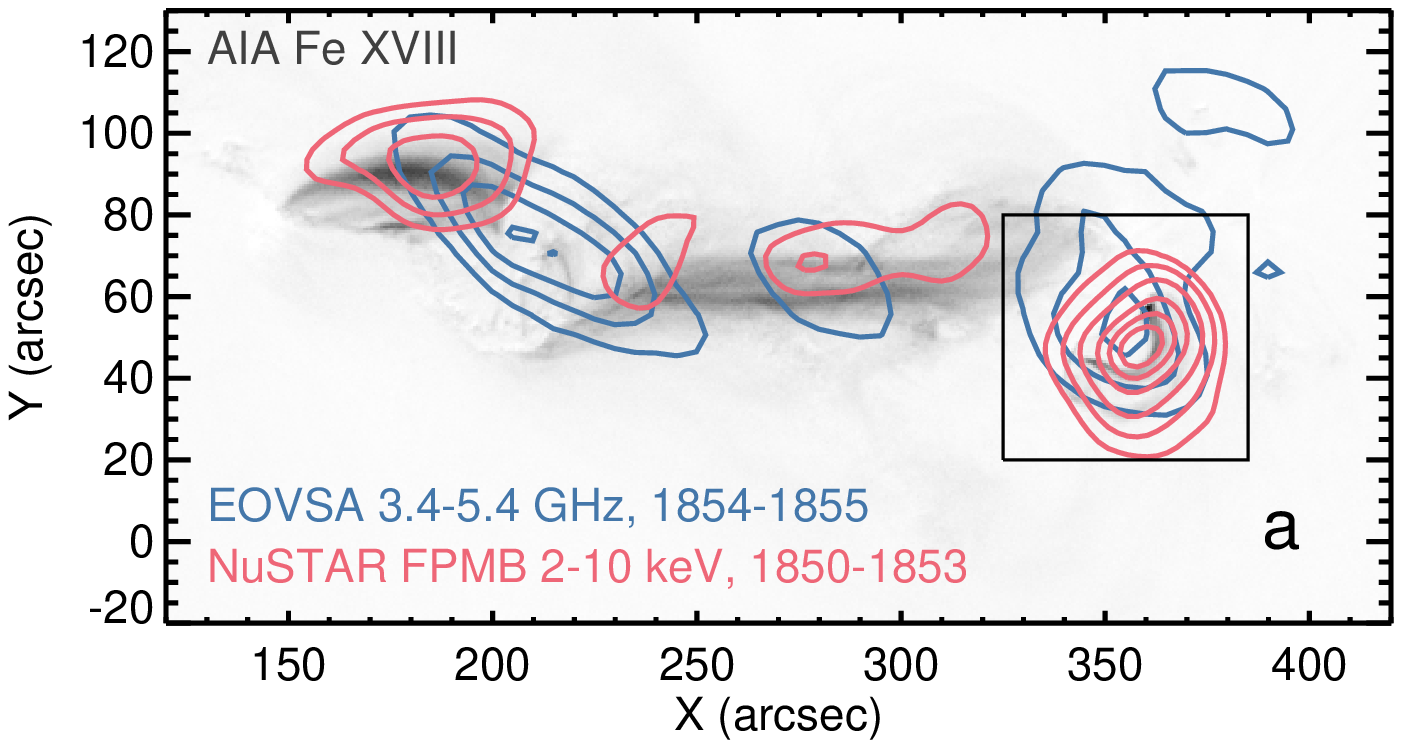}
	\includegraphics[width=0.32\linewidth]{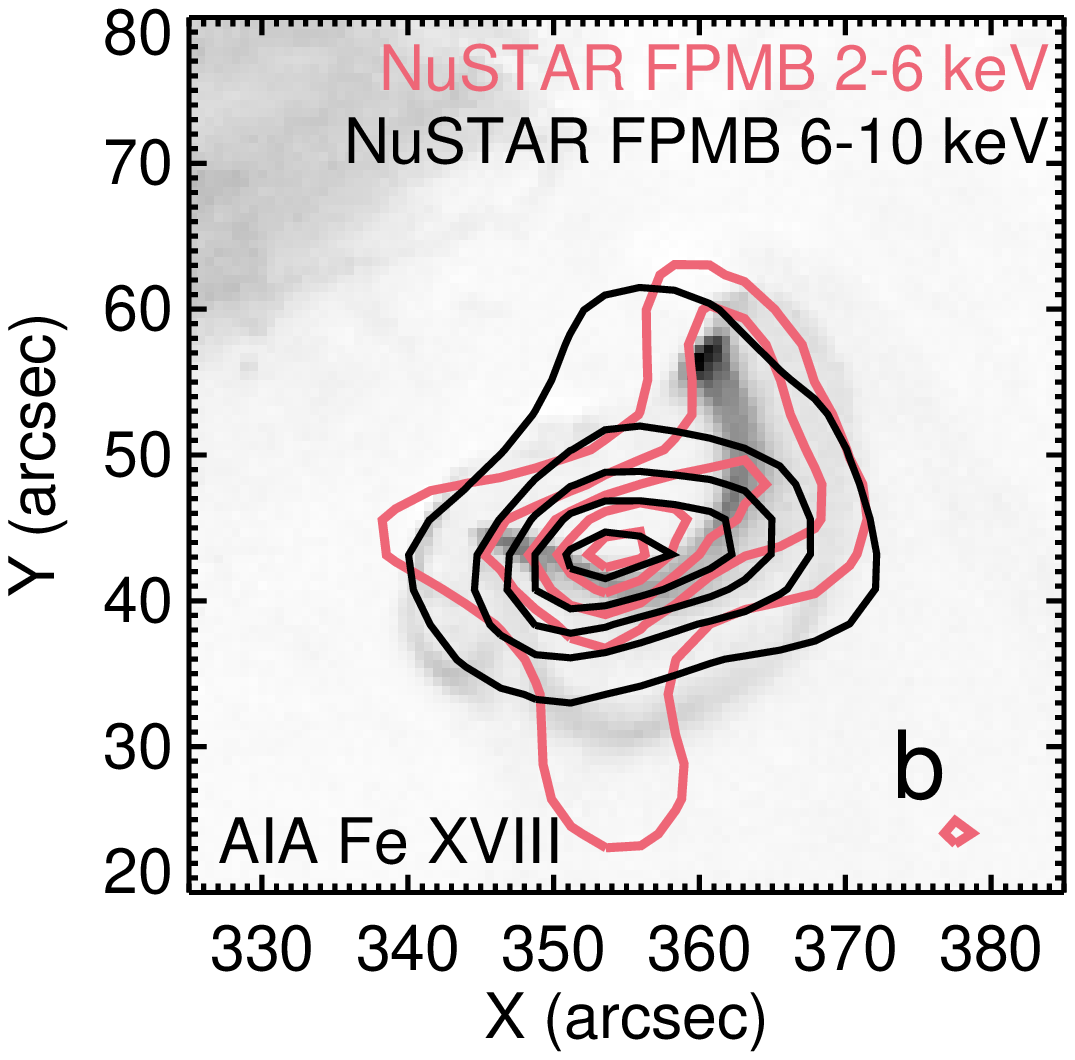}
	\includegraphics[width=0.32\linewidth]{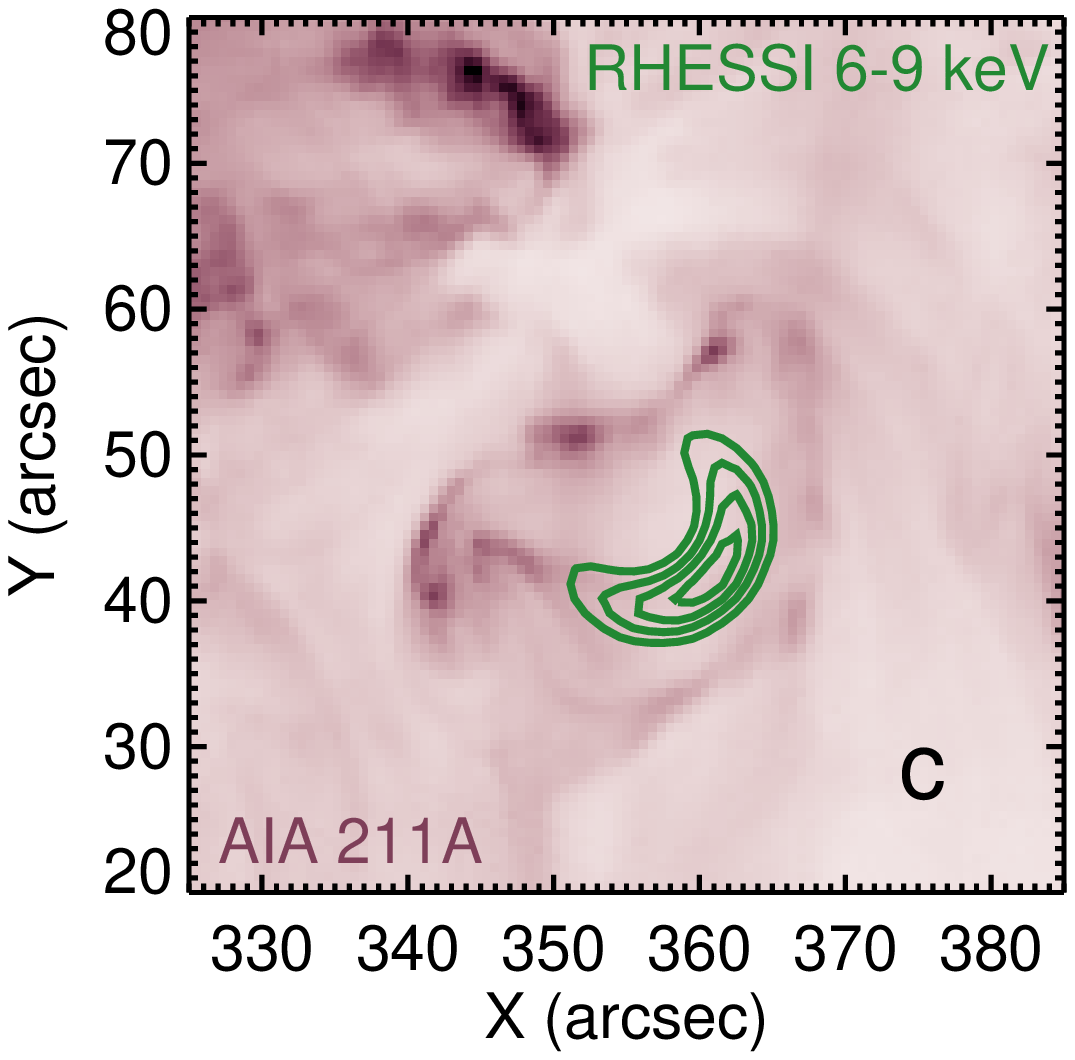}
	\includegraphics[width=0.267\linewidth, trim=1.8cm 0 0 0, clip=true]{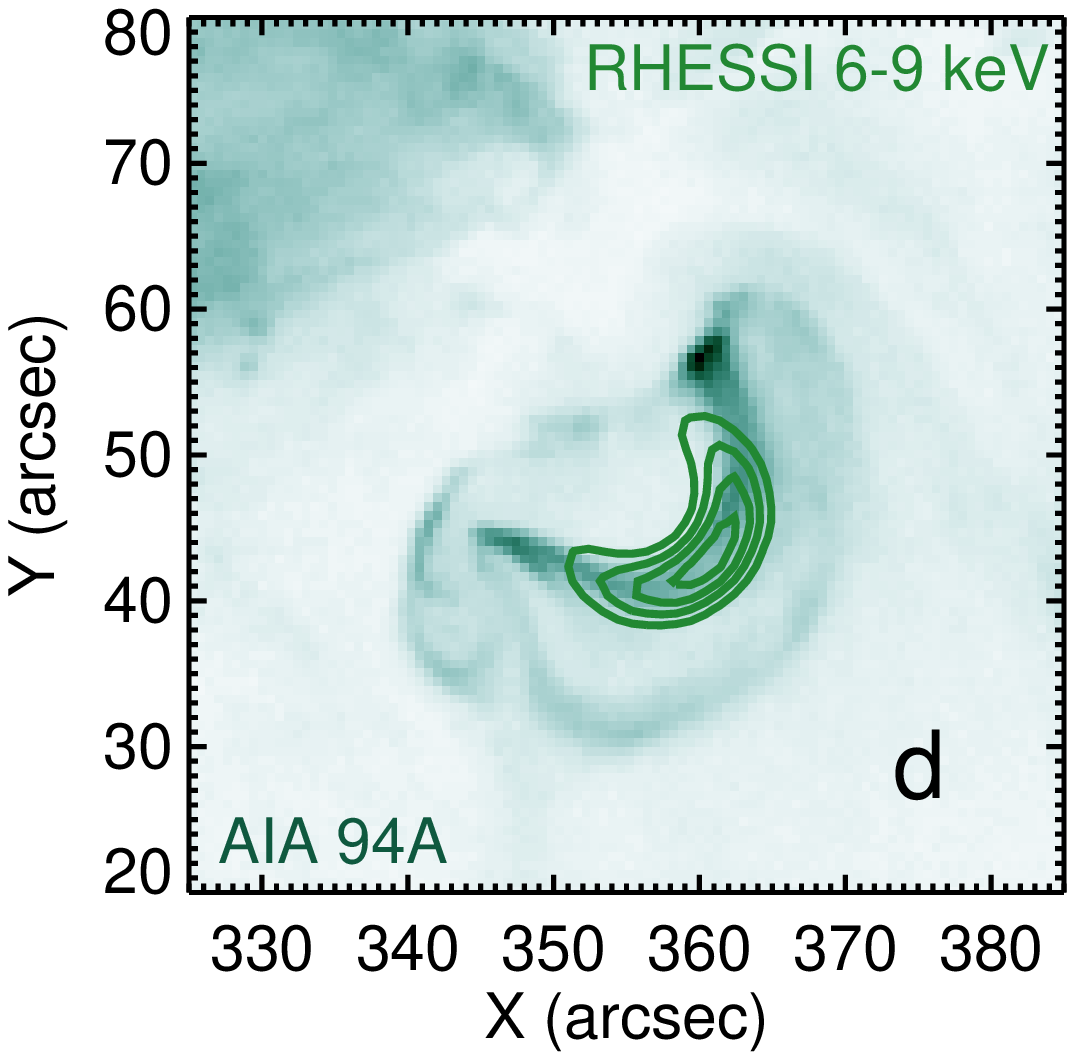}
	\includegraphics[width=0.267\linewidth, trim=1.8cm 0 0 0, clip=true]{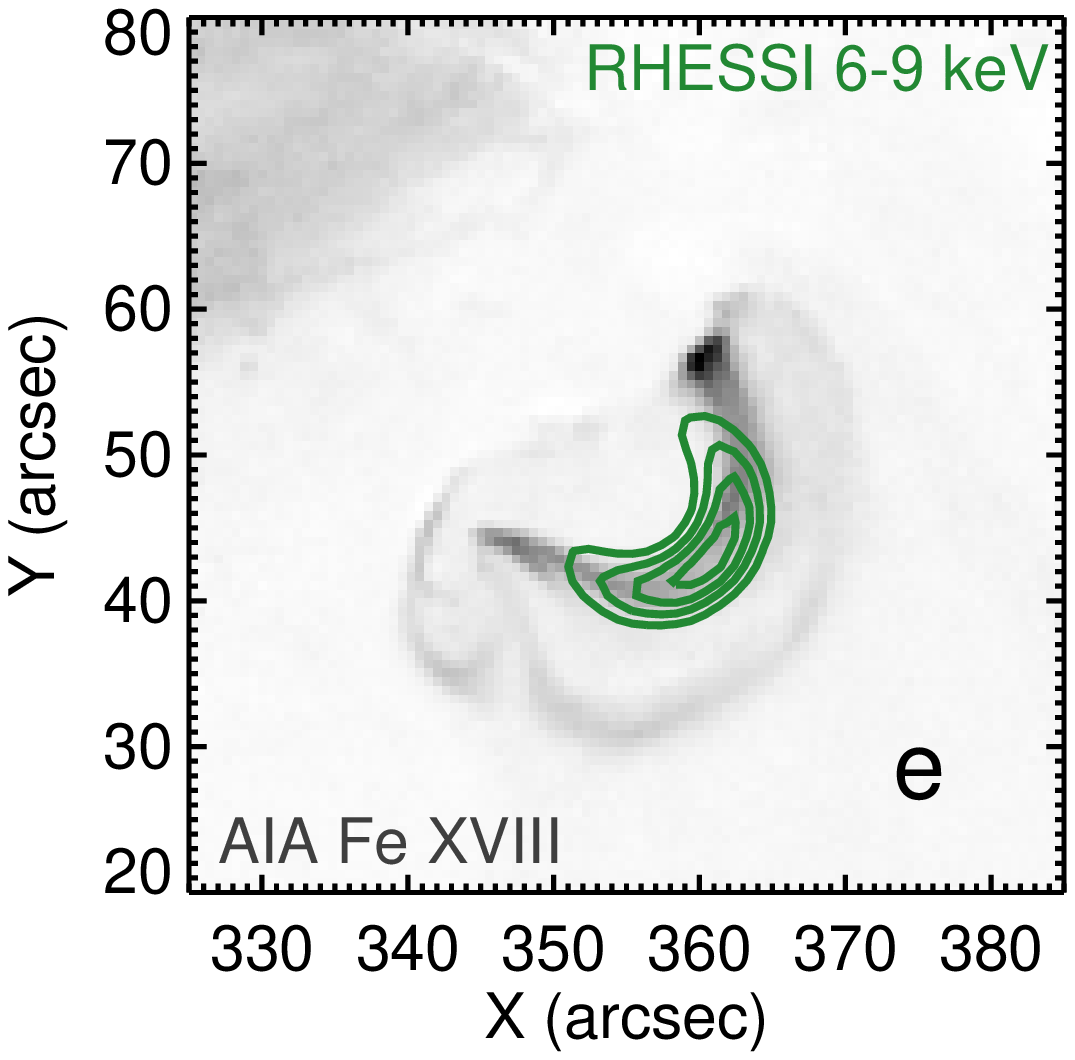}
\caption{ \textit{SDO}/AIA images overlaid with \nustar, \rhessi, and EOVSA emission.  All AIA images are integrated from 18:50 to 18:53 UT.  (Panel (a)) \nustar\ and EOVSA emission overlaid on an AIA image of the entire active region, from which the Fe XVIII component has been isolated.  The \nustar\ image has been deconvolved (50 iterations) and contour levels are 5, 10, 20, 40, 60, and 80\% of the maximum.  EOVSA contours are 30, 50, 70, and 90\% of the maximum.  The black box indicates the region shown in other panels.  (Panel (b)) \nustar\ emission deconvolved for 800 iterations (2-6 keV) and 100 iterations (6-10 keV) overlaid on an AIA image from the same time interval.  In Panels (a) and (b), \nustar\ images have been coaligned to AIA images, and \nustar\ data shown are from Focal Plane Module B only.  (Panels (c)-(e)) \rhessi\ \texttt{vis\_fwdfit} images overlaid on various AIA filter images.  (No coalignment is necessary.)  The HXR emission from both \rhessi\ and \nustar\  shows good agreement with the AIA flaring loop morphology.  
}
\label{fig:image}
\end{center}
\end{figure*}

Images of \nustar\ HXR emission are shown overlaid on AIA images in Figure \ref{fig:image}.  \nustar's two detector assemblies are termed Focal Plane Modules A and B (FPMA and FPMB).  For this event, FPMB recorded higher-quality data because some of the bright emission fell in the small gap between the detector quadrants of FPMA, so FPMB data are utilized for the images.  In Panel (a), FPMB data from 2 to 10 keV have been integrated over the three-minute interval indicated in Figure \ref{fig:lightcurves}, have had the instrument point spread function deconvolved for 50 iterations using the IDL procedure \texttt{max\_likelihood.pro}\footnote{Available within the IDL Astronomy User's Library at \url{https://idlastro.gsfc.nasa.gov}}, and have been coaligned to AIA data as previously described.  This figure also includes data from the Extended Owens Valley Solar Array (EOVSA), which is sensitive to microwave emission from flare-accelerated electrons \citep[e.g.][]{gary2018}.  Panel (b) shows a zoomed-in image of FPMB emission in two energy bands after 800 iterations of deconvolution for the 2-6 keV band and 100 iterations for the 6-10 keV band.  (Different iteration numbers are chosen based on the statistics available in each image.)  Both \nustar\ sources are shown at 10, 30, 50, 70, and 90\% level contours.  The \nustar\ and AIA source shapes are similar, and all HXR emission (in all available energy ranges) emanates primarily from the flaring loop(s).

The microflare was observed by \rhessi, although it was too faint for inclusion in the auto-generated \rhessi\ flare list.  At this late stage in \rhessi's mission (15.5 years post-launch), only detectors 1, 3, 6, and 8 were operating.  Analysis of a microflare this faint and at such low energies is challenging with \rhessi's performance at the time, but the array of spatial frequencies covered by this set of four subcollimators is sufficient to produce an image of the microflare using the \texttt{vis\_fwdfit} imaging method\footnote{See \url{https://hesperia.gsfc.nasa.gov/rhessi3/software/imaging-software/image-algorithm-summary/index.html} for a summary of \rhessi\ imaging algorithms.}, as was used in \citet{hannah2008}.  This method presupposes a source shape (in this case a loop) and forward-fits the source parameters to the observed visibilities.  The result of this method for energies 6--9 keV is shown in Panels (c)-(e) of Figure \ref{fig:image}.  \rhessi\ images produced using other imaging algorithms (e.g. \texttt{Clean}; not shown) and at higher energies (e.g. 8-10 keV) all show similar results; the HXR loop matches the position, loop shape, and rotation angle of the AIA loop.  \rhessi\ provides highly accurate source locations, and so no co-alignment with AIA was necessary.  In summary, all HXR observations, from both \nustar\ and \rhessi, reveal that HXR emission emanates primarily from the flaring coronal loop(s).

\begin{figure*}[htb]
\begin{center}
	\includegraphics[width=0.363\linewidth]{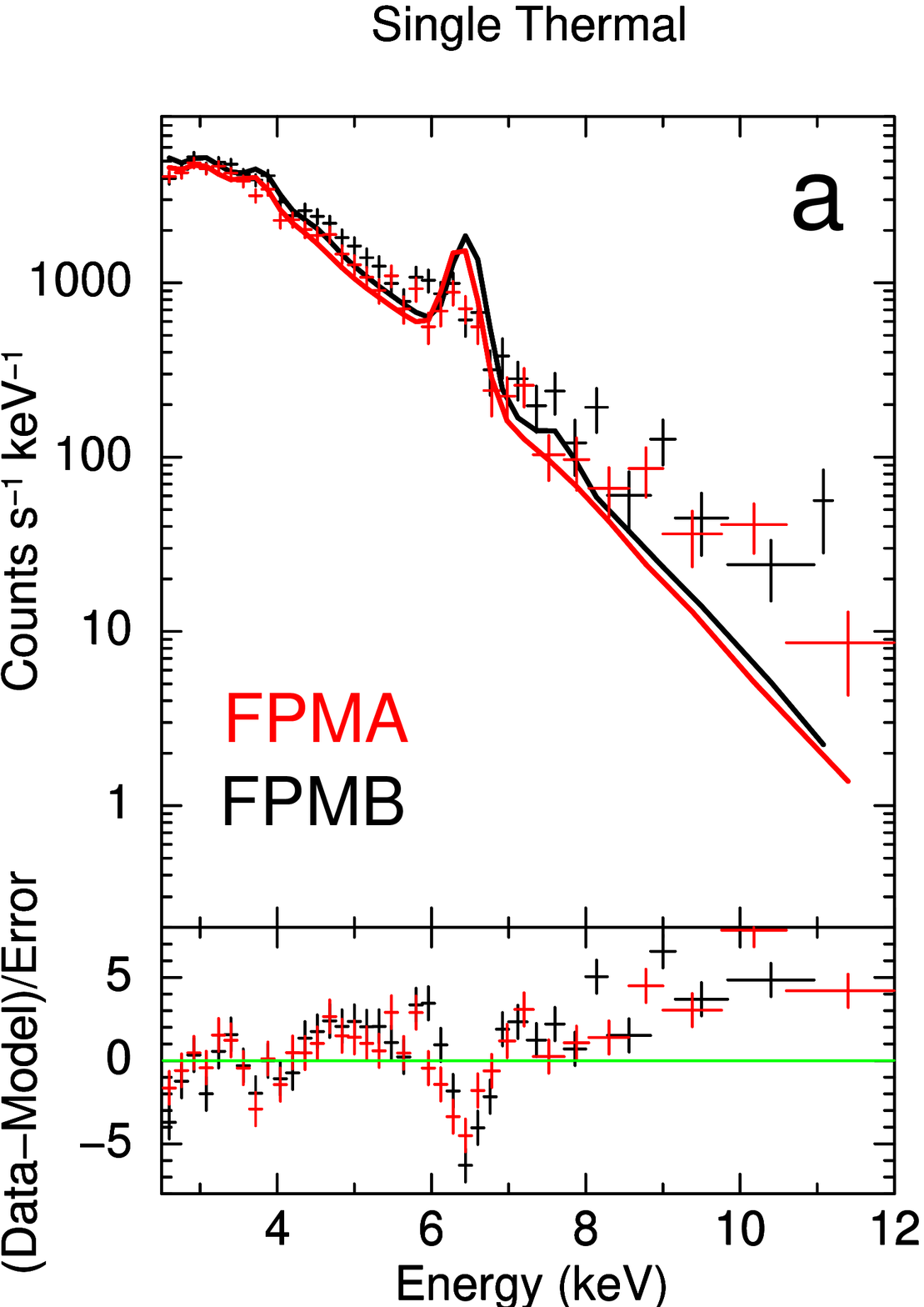}
	\includegraphics[width=0.30\linewidth, trim=2.6cm 0 0 0, clip=true]{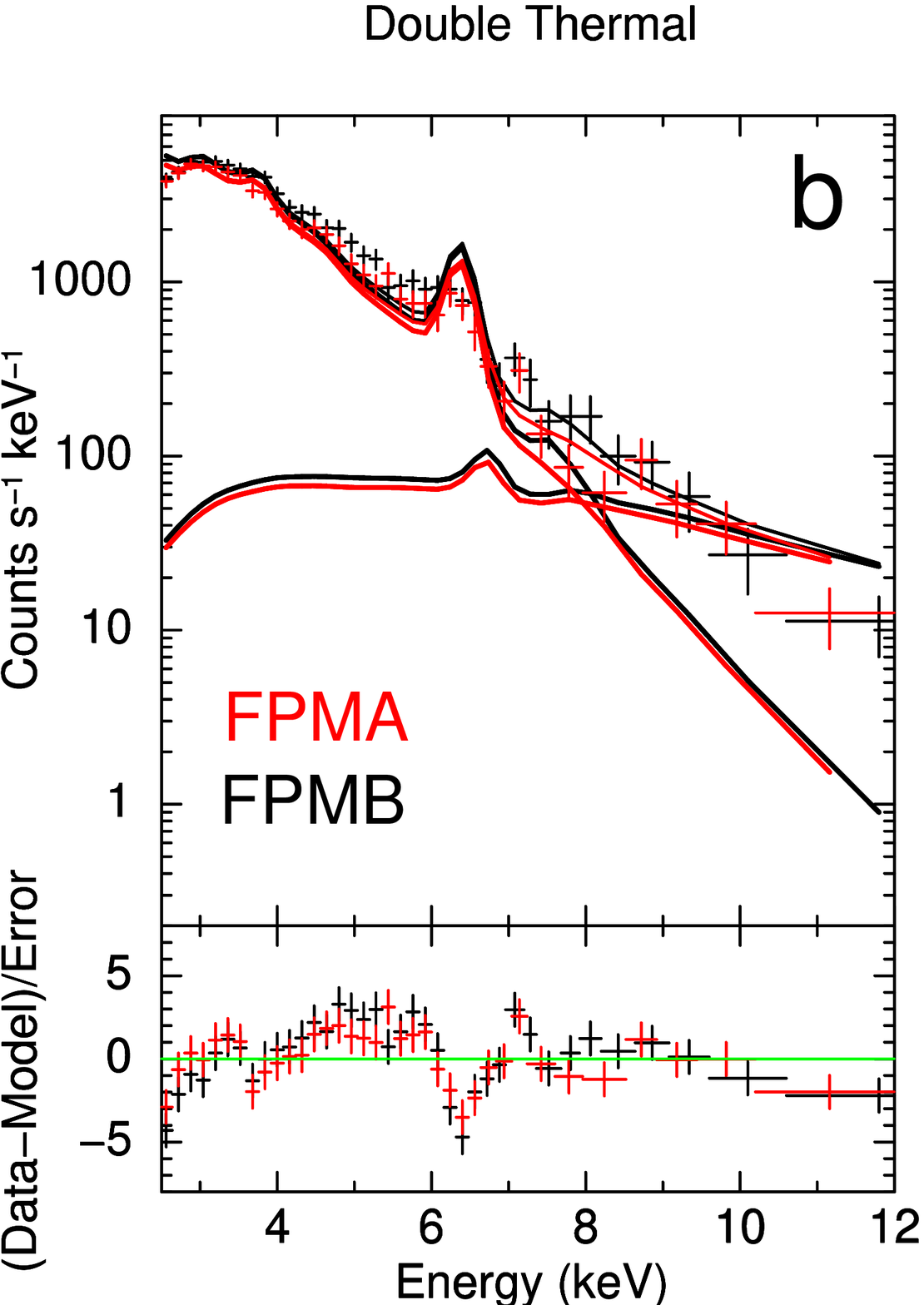}
	\includegraphics[width=0.30\linewidth, trim=2.6cm 0 0 0, clip=true]{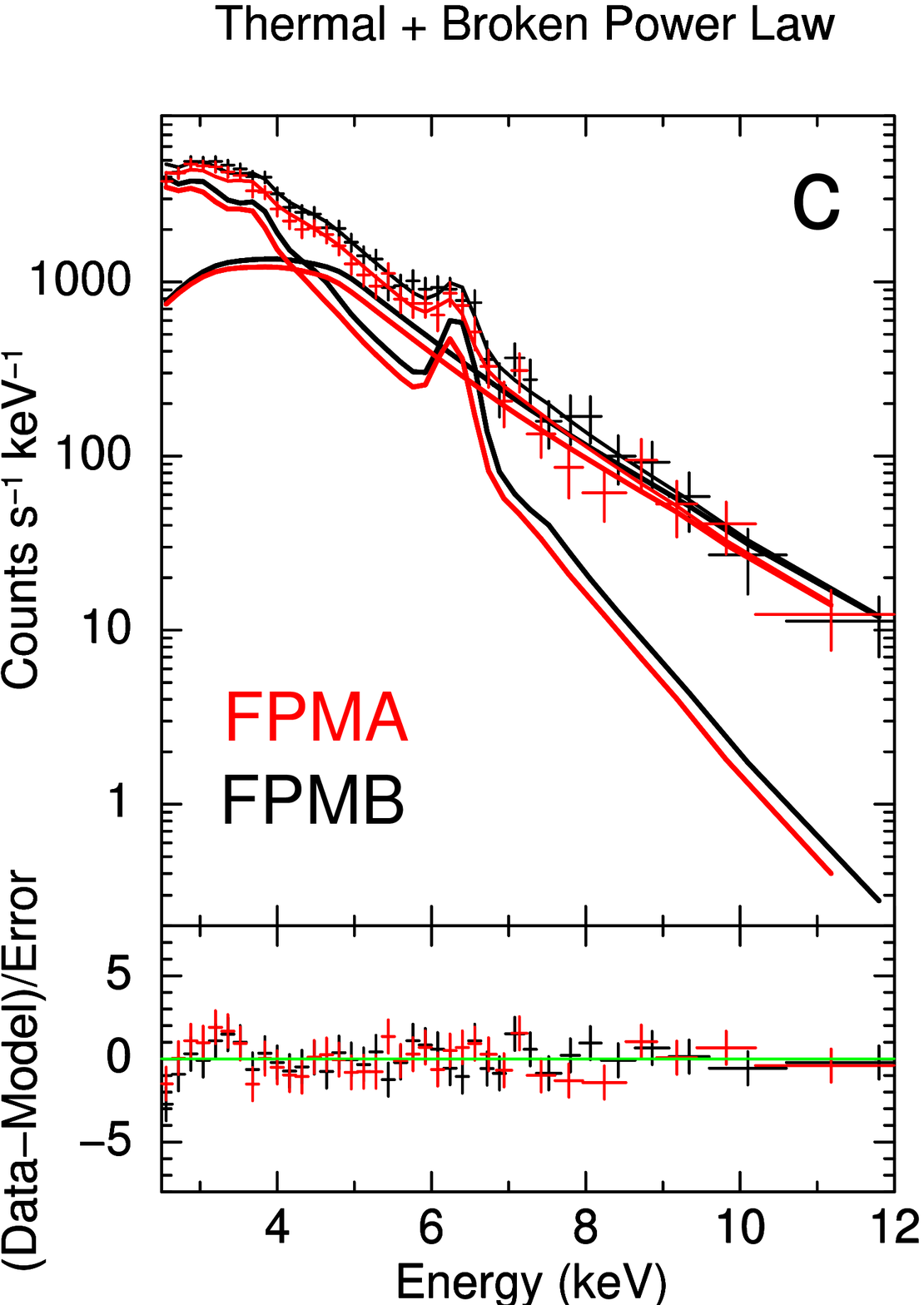}
\caption{ Spectral fits to \nustar\ data in the 18:50--18:53 UTC time interval for (a) an isothermal model, (b), a double-thermal model, and (c) an isothermal + broken power law model.  Count spectra shown are livetime- and pileup-corrected, and a gain correction has been included in the fit.  In each spectrum, the top panel shows the count spectrum (and fit) for the two telescopes (FPMA, red, and FPMB, black) and the bottom panel shows the error-normalized residual distribution.  The residuals indicate a more appropriate fit for the isothermal + broken power law model than for either of the purely thermal distributions, revealing a flare-accelerated electron distribution.  Fit parameter values are listed in Table \ref{tab:params}. }
\label{fig:xspec}
\end{center}
\end{figure*}

\begin{table*}[hp]
\begin{center}
\begin{threeparttable}
\caption{Fit parameters for all spectral models in Section \ref{sec:spectroscopy}, along with best fit statistic values.}
\def\arraystretch{1.5}
\begin{tabular}{|l|l|l|l|l|l|}
\hline			&	\multicolumn{3}{c|}{{XSPEC Models}} 						&	\multicolumn{2}{c|}{{OSPEX Models}}	\\
\hline
&	\multirow{2}{*}{\texttt{vapec}$^a$}	&	\multirow{2}{1.5cm}{\texttt{vapec$+$ vapec}$^a$} 	&	\multirow{2}{1.5cm}{\texttt{vapec$+$ bknpower}} 	
&	\multirow{2}{1.5cm}{\texttt{vth$+$ thick2}}		&	\multirow{2}{*}{\texttt{warm\_thick}}	\\ 
				&				&					&				&					&	\\
\hline
Temperature [MK] 			&12.6$^{+0.1}_{-0.3}$ & 	11.8$^{+0.4}_{-0.3}$	& 10.4$^{+0.4}_{-0.2}$ & 	10.3$^{+0.7}_{-0.7}$  	&	10.2$^{+0.7}_{-0.7}$	\\
EM [10$^{45}$ cm$^{-3}]$		& 4.0$^{+0.1}_{-0.1}$ & 	4.5$^{+0.1}_{-0.1}$	& 4.2$^{+0.5}_{-0.4}$ &	5.0$^{+1.3}_{-1.3}$ 	&	\\
Density [10$^9$ cm$^{-3}$] &			&					&				&					&	6.0$^{+2.0}_{-2.0}$ 	\\
\hline
Temperature$^b$ [MK]	&				&	400$^{+60}_{-180}$  	&				&					&	\\
EM [10$^{45}$ cm$^{-3}]$ 	&		&	0.015$^{+0.003}_{-0.003}$  &			&					&	\\
\hline
Break Energy [keV] 	&				&					& 5.0$^{+0.1}_{-0.1}$ &					&	\\
Photon Index$^c$ $\gamma$ &			&					& 5.5$^{+0.3}_{-0.2}$ &					&	\\
\multirow{2}{4cm}{Norm [phot keV$^{-1}$ cm$^{-2}$ s$^{-1}$ at 1 keV]}
				&				&					& \multirow{2}{*}{530$^{+60}_{-50}$} &					&	\\
				&				&					&				&					&     	\\
\hline
Cutoff Energy [keV] 	&				&					& 				&	6.2$^{+0.9}_{-0.9}$	&	6.5$^{+0.9}_{-0.9}$	 \\
Electron Index $\delta$ &				&					& 				&	6.2$^{+0.6}_{-0.6}$ 	&	6.3$^{+0.7}_{-0.7}$ 	\\
Electron Flux [10$^{35}$ e$^-$ s$^{-1}$] &&		& 				 &	2.1$^{+1.2}_{-1.2}$	&	1.8$^{+0.8}_{-0.8}$	\\
Loop Half Length [Mm]		  &				&					&				&					&     	15 (fixed)	\\
\hline
C-Statistic	 (reduced)		& 2.5			&	1.8				&	1.2			&					&	\\
$\chi^2$ value (reduced)	& 2.0			&	2.0				&	1.1			&	0.8				&	0.7	\\
\hline
\end{tabular}
 \begin{tablenotes}
\item [$^a$] The purely thermal fits are included for fit comparison purposes only; they are poorer fits than the thermal$+$non-thermal fits and are not good representations of this flare.
\item [$^b$] Parameter was allowed to vary only up to 40 keV (464 MK).
\item [$^c$] Index above the break. The index below the break was fixed at 2.
\end{tablenotes}
\label{tab:params}
\end{threeparttable}
\end{center}
\end{table*}%

\subsection{HXR spectral fitting}
\label{sec:spectroscopy}

An examination of the \nustar\ spectral data over the first three minutes of the flare indicated a small pileup component, and so a pileup correction was performed on the count spectra \citep[see Appendix C in][]{grefenstette2016}.  Since each \nustar\ event is assigned a ``grade,'' comparison of events of various grades gives an estimate of the pileup contribution, which can then be subtracted from the spectrum.  The necessary pileup correction is no more than 6.25\% in any 0.64 keV energy band.  Statistical uncertainties were widened to account for uncertainties in the subtracted components.  Additionally, a gain slope correction was included to account for variations in the \nustar\ gain that occur only at extremely high rates (e.g. livetime of a few percent) encountered at the Sun \citep{duncan2019}; this parameter was allowed to vary for XSPEC fits and was fixed to 0.95 (the best fit value) for OSPEX fits.  The gain adjustment resulted in a 5\% increase in temperature and a 20\% decrease in emission measure.

Following these pileup and gain corrections, fitting of various spectral models was performed using the XSPEC software package \citep{arnaud1996}.  Fits were performed simultaneously to FPMA and FPMB using the C-statistic to assess goodness of fit.  Figure \ref{fig:xspec} shows the results of fitting for three different spectral models: an isothermal model (\texttt{vapec}), a double thermal model (\texttt{vapec}+\texttt{vapec}), and an isothermal model plus a broken power law (\texttt{vapec}+\texttt{bknpower}).  The fit parameters are summarized in Table \ref{tab:params}.  For the \texttt{bknpower} component, the spectral index below the break (where emission is dominated by thermal plasma) was fixed at 2.  The simplest model (isothermal only) shown in Panel (a) exhibits a high-energy excess that is not well fit by the model.  The double thermal model shown in Panel (b) picks up this high-energy component but retains systematic residuals that are not well distributed across energy; the iron line intensity is poorly predicted.  Additionally, the temperature required by the hotter component (400 MK) is unreasonably high.  When the temperature was restricted to a more physical range (e.g. $<$100 MK), the fit value was always driven to the highest allowed temperature; the iron line was still poorly predicted; and the fit statistic values were worse.  The third model, which includes a broken power law in photon space arising from a non-thermal electron distribution, exhibits well distributed residuals and a significantly better fit statistic than the thermal-only fits.  This model is selected as the best fit to the data and reveals the presence of an accelerated electron distribution.

As is often the case in fitting HXR spectra, the power-law spectral break (which is related to the low-energy cutoff of the accelerated electron distribution) is poorly constrained in the presence of bright thermal plasma.  The fit value is best viewed as an upper limit; see Section 3 of \citet{holman2011} for a thorough discussion.  In fact, for this microflare, even a thermal component plus an \textit{unbroken} power law produced only a slight worsening of the fit, bringing the temperature up to $T=11.9\pm(0.9,0.6)$ MK and emission measure down to $EM=2.3\pm(0.3,0.6)\times 10^{45} $cm$^{-3}$.  Although we do not consider a single power law likely, since it would require an accelerated electron distribution extending down to extremely low energies, we use it to set one bound on the flare thermal parameters.  The resulting range of allowed parameters (for the thermal+power-law models) is $T=10.2-12.8$ MK and $EM=(1.7-4.7)\times 10^{45}$ cm$^{-3}$.  We consider this a range of allowed parameters, but the most likely ones are those in the last column of Table \ref{tab:params}.

AIA data were examined for consistency with the \nustar\ temperature.  A simple ratio was taken of data from AIA filters 131\AA\ and 94\AA\ (filters with significant and relatively isolated response to hot flare-temperature plasma).  This ratio yields temperatures of 9.2--10.5 MK between 18:50 and 18:53 UTC and an emission measure consistent with that obtained via the \nustar\ spectral fit.

\begin{figure*}[htb]
\begin{center}
	\includegraphics[width=0.35\linewidth, trim=0 0 0 0.66cm, clip=true]{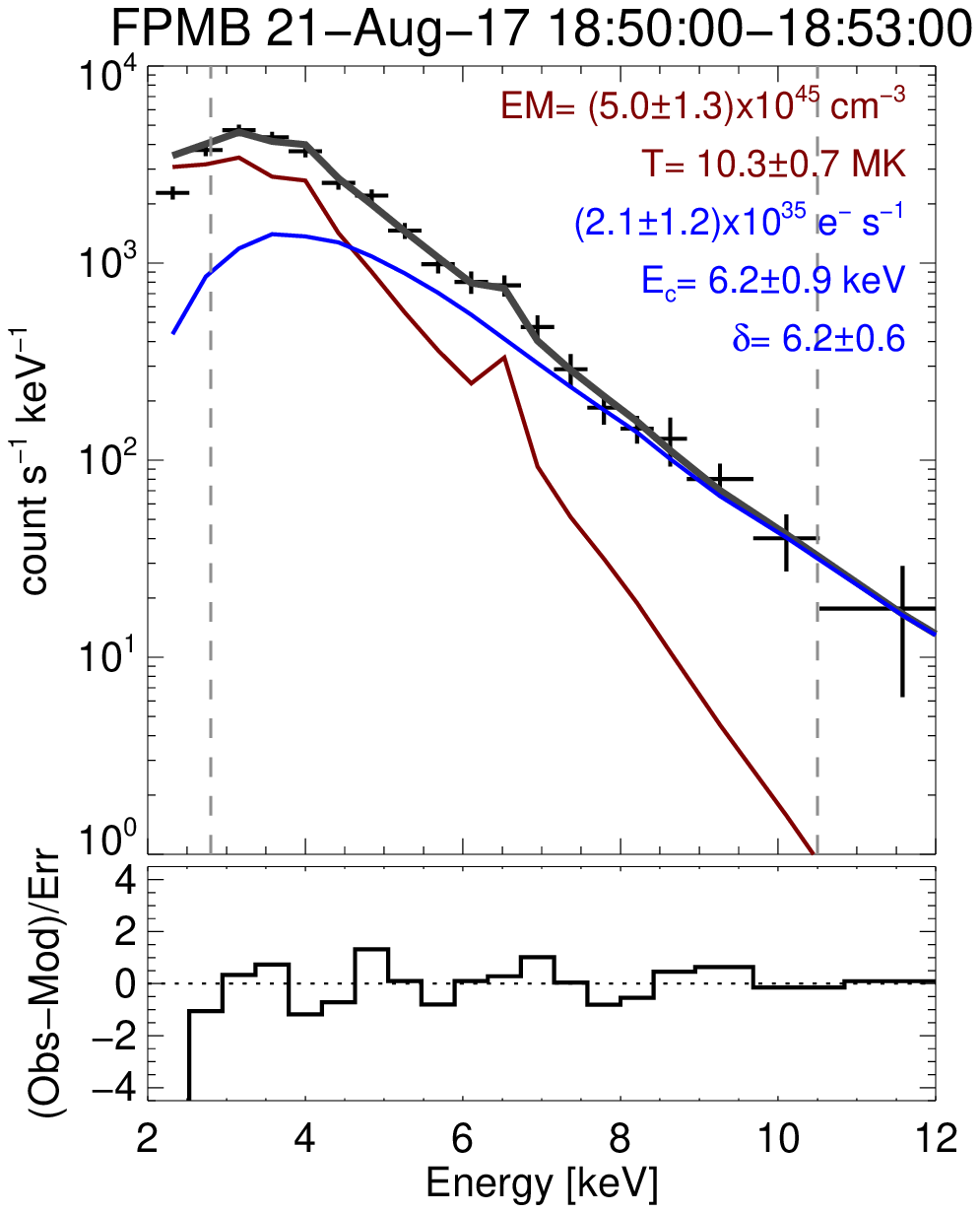}
	\includegraphics[width=0.35\linewidth, trim=0 0 0 0.66cm, clip=true]{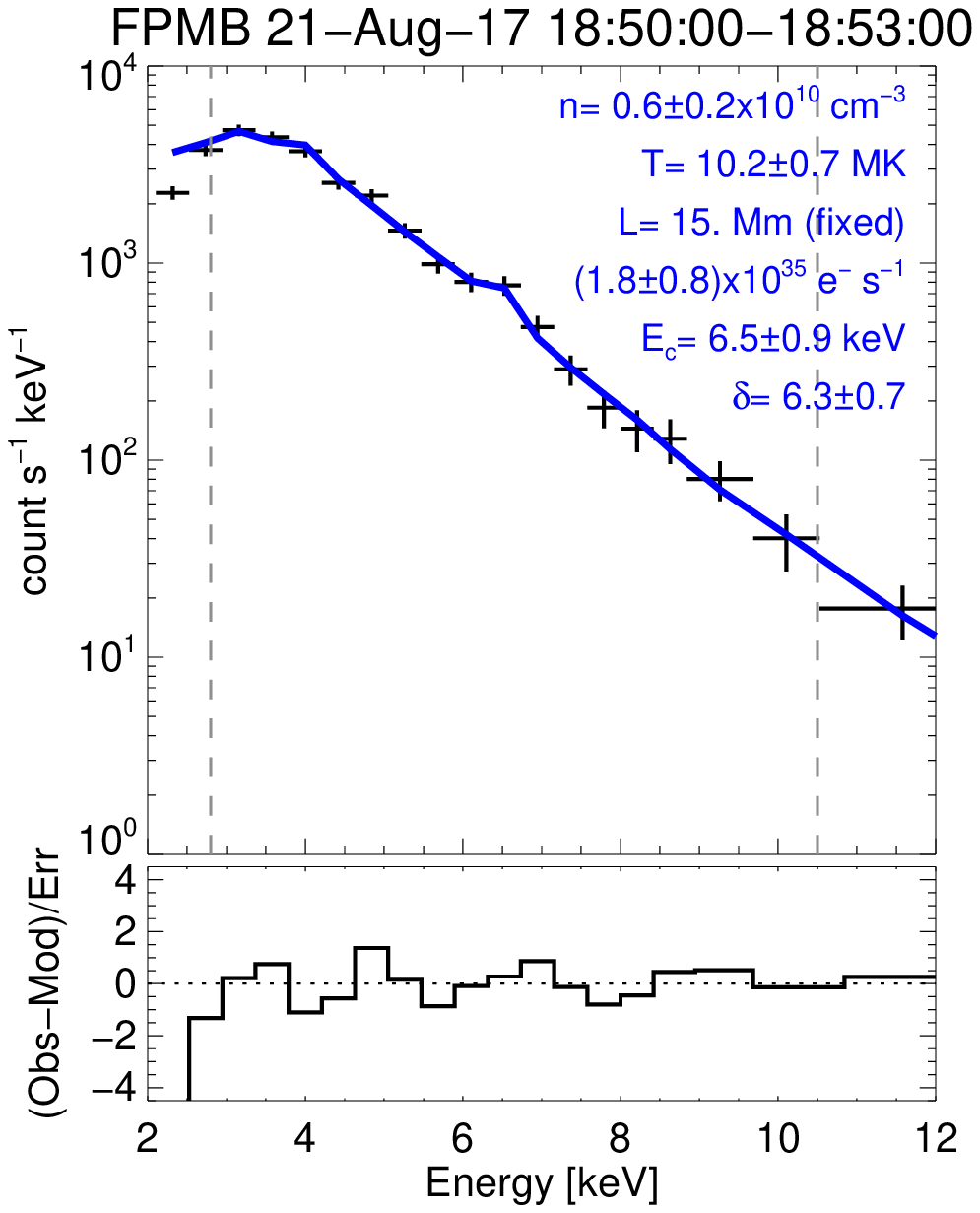}
\caption{Results of thick-target spectral fitting in OSPEX using models (left) \texttt{thick2} and (right) \texttt{thick\_warm}, which model an accelerated electron distribution propagating in a cold or warm plasma target, respectively.  Fits were performed to FPMB data only.  The warm target model fits the data well with no additional thermal component needed, indicating that the thermal plasma arises from energetic electron thermalization within the loop.  For the warm-target model, the loop half length was fixed to 15 Mm from AIA images and both temperature and density were allowed to vary.
}
\label{fig:ospex}
\end{center}
\end{figure*}

A more physical model than a broken power law is to directly fit an electron distribution to the X-ray data.  To accomplish this, a thick-target model of X-rays emitted by accelerated electrons (\texttt{thick2}) was fit to \nustar\ data along with an isothermal component (\texttt{vth}) in the spectral fitting package OSPEX\footnote{See \url{https://hesperia.gsfc.nasa.gov/rhessi3/software/spectroscopy/spectral-analysis-software/index.html}.}, which is commonly used to fit solar HXR flares.  Since the emission is integrated over the spatial extent of the flare and over a few minutes, we assume that the energetic electron distribution must completely thermalize, so no thin-target fit was performed.  Because OSPEX performs only single-instrument fits and only uses a chi-squared value as a fit parameter, data from FPMB (which has better coverage of this flare) were selected and rebinned to ensure at least 10 counts in each energy bin.  The result of this fitting is shown in Panel (a) of Figure \ref{fig:ospex}.  The parameters obtained from this fit are shown in Table \ref{tab:params} and are consistent with the parameters obtained from the broken power law fit in Figure \ref{fig:xspec}.  (Note that in a thick-target model, the electron and photon spectral indices $\delta$ and $\gamma$ are related by $\delta=\gamma+1$.)  An energy-integrated electron rate of $(2.1\pm 1.2)\times 10^{35}$ electrons s$^{-1}$ is obtained, along with a low-energy cutoff of $6.2\pm 0.9$ keV and an electron power-law index of $6.2\pm 0.6$.

The \texttt{thick2} model assumes a cold plasma target, requiring any thermal plasma to be added as a separate component.  On the other hand, the warm thick-target model (WTTM) implemented in OSPEX as \texttt{thick\_warm} provides a self-consistent fit assuming the presence of a thermal component due to electrons thermalized out of the accelerated population \citep[see][]{kontar2015, kontar2019}.  We find that the \texttt{thick\_warm} model can well fit the entire \nustar\ spectrum for this flare, with no additional thermal component required; see panel (b) of Figure \ref{fig:ospex}.  For this fit, the warm plasma length was fixed to 15 Mm (the loop half-length from AIA images) and both temperature and density were allowed to vary, as were the spectral index, low-energy cutoff, and normalization of the accelerated electron distribution\footnote{We note that our fitting approach differs from that suggested by \citet{kontar2019} because precision on the low energy cutoff is not our goal.}.  The obtained results are almost identical to those obtained via the cold thick-target model (CTTM), indicating that the CTTM is a valid approximation for this case.

In summary, a multitude of spectral fitting tools and models were utilized to fit the observed \nustar\ spectrum, including the X-ray fitting tools commonly used by both the solar and non-solar astrophysics communities.  All of these tools and models point toward a thermal component plus a non-thermal, flare-accelerated electron population, and the different methods produce consistent quantitative results.  The gain and pileup corrections mentioned at the start of this section had small impacts on the fit parameters but did not have any impact on which model provided the best fit.  The next section will interpret the spectral fitting results for their consequences on the dynamic evolution of this flare.

\section{Discussion}

\nustar\ spectral fitting indicates that the non-thermal bremsstrahlung X-ray flux intensity is greater than or equal to the thermal flux intensity above \app4.5 keV (see Figures \ref{fig:xspec} and \ref{fig:ospex}).  Further evidence for the non-thermal, rather than thermal, nature of the emission comes from the close relationship between the \nustar\ high-energy emission and the derivative of the \goes\ emission (see Figure \ref{fig:lightcurves}, Panel (g)).  This relationship, known as the Neupert effect, is commonly observed in larger flares and is often interpreted to indicate that the energy collisionally deposited by non-thermal electrons is responsible for the generation of the observed thermal plasma \citep{neupert1968, veronig2002}.

 \nustar\ does not have fine enough angular resolution to resolve much spatial detail of the flare source shape at high energies, but the 6--9 keV \rhessi\ source map in Figure \ref{fig:image} shows a loop matching the location, elongation, and orientation of the loop observed by AIA.  Therefore, the non-thermal emission observed by \nustar\ and \rhessi\ emanates from the flare loop, \textit{not} from its footpoints, akin to the ``coronal thick target'' flares studied by \citet{veronig2004, veronig2005, fleishman2016}.  This differs from the standard thick-target flare model in which energy is transported from the corona to the chromosphere primarily by accelerated electron beams, which emit strong bremsstrahlung radiation at their thermalization locations in the chromosphere and much fainter HXR emission from collisions in the tenuous corona.

A look at the energetics supports the coronal thick target scenario.  The non-thermal fit to \nustar\ data utilizing the warm-target model (described in Section \ref{sec:spectroscopy}) reveals that the accelerated electron distribution extends down to a cutoff energy of \app$6.5$ keV, or possibly even lower given that the cutoff is not well constrained in the particular way we are utilizing \texttt{thick\_warm}, and has a spectral index of \app$6.3$. This distribution has an average electron energy of \app8 keV.  To investigate the plasma within which this distribution propagates, we made geometric estimates from AIA.  We measured the area of the Fe XVIII AIA loop, using two thresholds on pixel brightness (one liberal and one conservative).  Taking the loop to be a bent cylinder with (line-of-sight) depth equal to its (plane-of-sky) width, the estimated volume is (1.2--2.2)$\times 10^{26}$ cm$^3$.  The full loop length, measured directly from Fe XVIII AIA images, is 24--32 Mm.  Assuming that volume is filled with the thermal plasma indicated by the cold-target fit, and assuming a filling factor of unity, this leads to a loop density of (2.8--6.3)$\times 10^9$ cm$^{-3}$.  No projection effects have been corrected for in these estimates, and if the filling factor is less than unity, then the density is even higher.  However, we note the calculated density is consistent with that fit in the warm-target model ($6\times10^9$ cm$^{-3}$), showing good correspondence between the two models.

If the flare-accelerated electrons are injected into the very top of the loop and must travel one half loop length to reach the chromosphere, the column density encountered is high enough to collisionally thermalize all electrons below 5.4--7.1 keV.  If the electrons are mirrored by the relatively stronger magnetic fields encountered at lower altitudes, then they encounter even higher column density.  Most electrons below 9.4--12.4 keV would be collisionally stopped by even 1.5 bounces in the loop.  Therefore, it is reasonable that the observed accelerated electron distribution deposits most of its energy via collisions in the corona.  This is  different from the expected behavior in larger-energy flares, in which electrons are observed up to many tens or hundreds of keV, and low-energy electron cutoff energies are often tens of keV; those distributions would easily penetrate loops of typical coronal densities and produce the bright HXR footpoints typically observed in large \rhessi\ flares.  We note that even in larger flares, densities can sometimes be high enough to produce this effect, as in the \citet{veronig2004} flares.

From the spectral fits, the non-thermal electrons in this flare deposit energy at a rate of \app$2\times10^{27}$ erg/s, for a total of \app$4 \times 10^{29}$ ergs when integrated over the three minutes chosen for spectroscopy.  The average thermal energy over this time is estimated from \nustar\ spectroscopy to be \app$3\times 10^{27}$ erg; these and other energetics values are summarized in Table \ref{tab:param}.  We have compared this to thermal energy estimates for this flare made using \textit{SDO}/AIA data.  Using the differential emission measure estimation method of \citet{cheung2015}, we find that the thermal energy in the flaring pixels rises from 1.3$\times 10^{27}$ to 2.4$\times 10^{28}$ erg between 18:50 and 18:53 UT, reaching a peak at 2.7$\times 10^{28}$ at 18:54 UT when considering only the emission measure above 8 MK (corresponding to the plasma measured by \nustar).  When considering all temperatures, the thermal energy peaks at 3.6$\times 10^{28}$ erg.  This means that the ratio of non-thermal electron energy (over the three minutes chosen for analysis) to maximum observed thermal energy (occurring shortly after this interval) is a factor of 10.  This is a larger non-thermal ratio than any of the C class flares studied using \rhessi\ and \goes\ data by \citet{warmuth2016}, but is a very typical ratio for the larger flares in that study (see, e.g., Figure 7 of that work).  \citet{aschwanden2016} found an average non-thermal-to-thermal energy ratio of 0.15-6.7, depending on the method used for the non-thermal calculation.  (That study relied on two different ways to estimate the non-thermal electron energy.  Their particular application of the warm target method resulted in artificially high non-thermal energy estimates as explained in \citet{kontar2019}, but their other, ``cross-over'' method provided a conservative lower limit on the non-thermal energies, so the true ratio probably lies in between.)  

\rhessi\ flares of this thermal energy are close to \rhessi's sensitivity limit; this is evident from Figures 13 and 18 of \citet{hannah2008}. Figure 18 shows the rollover in the \rhessi\ flare frequency distribution below \app$10^{29}$ erg, and Figure 13 shows that this flare, with a temperature of \app10 MK and an EM of \app$5\times 10^{45}$ cm$^{-3}$, would not have been included in that previous study. The reason for this is that the selection criteria used in that study to produce reliable automated fits would have excluded this microflare. So future work could revisit the \rhessi\ data to search for similar microflares, during times the instrument was performing optimally. However, it might be that with \nustar's sensitivity to faint emission, flares with higher non-thermal-to-thermal energy ratios can be observed than in the past.  This implies that the ratios of non-thermal to thermal energies in \citet{warmuth2016} might be much more consistent across flare size if high HXR sensitivity were available for more small flares.  Due to \nustar's limited solar observing time and limited throughput (which restricts its spectral dynamic range), a thorough study of this point must wait for a future, space-based direct HXR imager optimized for the Sun.

\begin{table}[htp]
\caption{{Energetics values}}
\begin{tabular}{|l|l|}
\hline
Spectral index	&	6.3 \\
Low energy cutoff	&	6.5 keV \\
Average non-thermal power,	&	$2\times 10^{27}$ erg s$^{-1}$ \\
\hspace{0.1cm} 18:50--18:53 & \\
Energy deposited over 3 min	&	 $4 \times 10^{29}$ erg	\\ 
\nustar\ thermal energy &  	$3 \times 10^{27}$ erg	\\
 \hspace{0.1cm} (isothermal, avgd 18:50--18:53)		& \\
 AIA thermal energy & 	$4 \times 10^{28}$ erg \\
\hspace{0.1cm} (DEM, peak energy, 18:54)	& \\
 Number accelerated electrons 	&	$2 \times 10^{35}$ e$^-$ s$^{-1}$		\\
Average electron energy		&	8 keV	\\
\hline
\end{tabular}
\label{tab:param}
\end{table}%

The coronal thick target scenario does not rule out the occurrence of chromospheric evaporation.  The higher-energy tail of the accelerated electron distribution could persist through enough column length to precipitate to the flare footpoints and deposit energy to inspire evaporation.  Alternatively, footpoint heating via conduction could serve the same function; \citet{warmuth2016} found conductive loss rates to be dominant in small flares.  Here, the radiative loss rate assuming an average temperature of 10 MK and an emission measure of $5\times10^{45}$ cm$^{-3}$ is $5\times10^{23}$ erg s$^{-1}$ and the conductive loss rate assuming Spitzer conductivity is approximately $10^{26}$ erg s$^{-1}$.  However, it is difficult to assess the footpoint area for this flare, which has a strong impact on the conductive losses.  If the scaling relations in \citet{warmuth2016} extend down (in temperature) to this flare, then the conductive loss rate would be \app$5\times 10^{26}$ erg s$^{-1}$, or about 0.25 of the non-thermal power, a significant fraction.  The scenario here is similar to the flares studied by \citet{veronig2004}, which first posited coronal thick-target flares; in those cases both electron precipitation and conduction contributed to chromospheric evaporation.  The two flares described in \citet{veronig2004} also exhibited gradual behavior (rather than impulsive) and rather steep electron power-law distributions, as does this microflare.

Microflare loop lengths are not, in general, smaller than regular flare loop lengths \citep{hannah2008, glesener2017}, and microflares tend to have lower low-energy cutoffs (although this could be an observational bias) and steeper spectra than larger flares do; this microflare observation continues that trend.  Given these two qualities, it is expected that flares smaller in energy should exhibit, on average, a greater degree of coronal thick-target behavior than larger flares do, and this trend should continue to even fainter flares.  It is likely appropriate to model such flares with a collisional deposition of energy throughout the flare loop rather than only at the footpoints.

The \nustar\ spectral observations clearly show that the measured \rhessi\ photons for this flare mainly emanate from a non-thermal source. This challenges the usual assumption made in HXR microflare analysis that the lowest energy ($\lesssim$7 KeV) counts are dominated by thermal emission. Further studies are needed to investigate how commonly flares with non-thermal emission down past the iron complex energies occur; the iron complex will be an important disambiguator in those studies, as it is here.  For this purpose, it will also be useful to exploit the database of AIA observations in tandem with the \rhessi\ archive, as was shown to be necessary in \citet{inglis2014, ryan2014}; and \citet{aschwanden2015}.  As the derived non-thermal energy content depends strongly on the low energy cutoff, a fresh look at the \rhessi\ microflare statistics with different spectral assumptions could potentially provide a significant update to microflare energetics.

\section{Summary}

In summary, we have reported the first evidence of non-thermal HXR emission from the Sun using direct focusing instruments.  This non-thermal emission was observed in a small, A5.7 class microflare.  The flare-accelerated electrons have an average energy of \app8 keV and extend down to a low-energy cutoff energy of \app6.5 keV, dominating the X-ray spectrum down to $<$5 keV.  A clear Neupert effect is observed.  The non-thermal electrons deposit most of their energy in the coronal loop, unlike most larger flares, which deposit collisional energy primarily at the footpoints.  The observation confirms that flare particle acceleration occurs even in the faintest flares observable with today's instrumentation, and that the non-thermal energies can be large in comparison to previous observations of small flares that were studied with less sensitive instruments.  Based on physical arguments, we suggest that extremely small microflares and nanoflares may be likely to be coronal thick-target flares, and simulations of such flares would be best served by depositing flare-accelerated electron energy throughout the corona and chromosphere, not only at footpoints.

Future direct-focusing HXR instrumentation with greater sensitivity, solar optimization, and much more solar observing time will allow the measurement of many more microflares like these, as well as fainter ones.  Such instrumentation has been developed with the \foxsi\ sounding rocket program and has been proposed as the \textit{Fundamentals of Impulsive Energy Release in the Corona Explorer (FIERCE)} spacecraft concept \citep{shih2019}.  Instruments based on this technology will allow full exploration of how particle acceleration scales to small flares, how frequently small flares occur, and how capable they are of heating the corona.

\acknowledgments

Support for this work was provided by an NSF Faculty Development Grant (AGS-1429512), an NSF CAREER award (NSF-AGS-1752268), and the NASA NuSTAR Guest Observer program (80NSSC18K1744).  IGH is supported by a Royal Society University Fellowship.  The authors are indebted to Natasha Jeffrey and Alexander Warmuth for their perspectives on this work.  Some figures were produced using IDL color-blind-friendly color tables by \citet{pjwright}.


\begin{thebibliography}

\bibitem[Arnaud(1996)]{arnaud1996} Arnaud, K.~A.\ 1996, Astronomical Data Analysis Software and Systems V, 17

\bibitem[Aschwanden et al.(2015)]{aschwanden2015} Aschwanden, M.~J., Boerner, P., Ryan, D., et al.\ 2015, \apj, 802, 53

\bibitem[Aschwanden et al.(2016)]{aschwanden2016} Aschwanden, M.~J., Holman, G., O'Flannagain, A., et al.\ 2016, \apj, 832, 27

\bibitem[Athiray et al.(2020)]{athiray2020} Athiray, P.S., et al.\ 2020,  \apj, accepted

\bibitem[Cheung et al.(2015)]{cheung2015} Cheung, M.~C.~M., Boerner, P., Schrijver, C.~J., et al.\ 2015, \apj, 807, 143

\bibitem[Christe et al.(2008)]{christe2008} Christe, S., Hannah, I.~G., Krucker, S., McTiernan, J., \& Lin, R.~P.\ 2008, \apj, 677, 1385-1394 

\bibitem[Dahlin et al.(2016)]{dahlin2016} Dahlin, J.~T., Drake, J.~F., \& Swisdak, M.\ 2016, Physics of Plasmas, 23, 120704

\bibitem[Dahlin et al.(2017)]{dahlin2017} Dahlin, J.~T., Drake, J.~F., \& Swisdak, M.\ 2017, Physics of Plasmas, 24, 092110 

\bibitem[Del Zanna(2013)]{delzanna2013} Del Zanna, G.\ 2013, \aap, 558, A73 

\bibitem[Duncan et al.(2019)]{duncan2019} Duncan, J.~M., Glesener, L., Hannah, I., et al.\ 2019, American Astronomical Society Meeting Abstracts, 234, 204.04 

\bibitem[Fleishman et al.(2016)]{fleishman2016} Fleishman, G.~D., Xu, Y., Nita, G.~N., et al.\ 2016, \apj, 816, 62

\bibitem[Gary et al.(2018)]{gary2018} Gary, D.~E., Chen, B., Dennis, B.~R., et al.\ 2018, \apj, 863, 83 

\bibitem[Glesener et al.(2016)]{glesener2016} Glesener, L., Krucker, S., Christe, S., et al.\ 2016, \procspie, 9905, 99050E 

\bibitem[Glesener et al.(2017)]{glesener2017} Glesener, L., Krucker, S., Hannah, I.~G., et al.\ 2017, \apj, 845, 122 

\bibitem[Grefenstette et al.(2016)]{grefenstette2016} Grefenstette, B.~W., Glesener, L., Krucker, S., et al.\ 2016, \apj, 826, 20 

\bibitem[Hannah et al.(2008)]{hannah2008} Hannah, I.~G., Christe, S., Krucker, S., et al.\ 2008, \apj, 677, 704 

\bibitem[Hannah et al.(2011)]{hannah2011} Hannah, I.~G., Hudson, H.~S., Battaglia, M., et al.\ 2011, \ssr, 159, 263

\bibitem[Hannah et al.(2019)]{hannah2019} Hannah, I.~G., Kleint, L., Krucker, S., et al.\ 2019, \apj, 881, 109

\bibitem[Harrison et al.(2013)]{harrison2013} Harrison, F.~A., Craig, W.~W., Christensen, F.~E., et al.\ 2013, \apj, 770, 103 

\bibitem[Holman et al.(2011)]{holman2011} Holman, G.~D., Aschwanden, M.~J., Aurass, H., et al.\ 2011, \ssr, 159, 107 

\bibitem[Inglis, \& Christe(2014)]{inglis2014} Inglis, A.~R., \& Christe, S.\ 2014, \apj, 789, 116

\bibitem[Kontar et al.(2015)]{kontar2015} Kontar, E.~P., Jeffrey, N.~L.~S., Emslie, A.~G., \& Bian, N.~H.\ 2015, \apj, 809, 35 

\bibitem[Kontar et al.(2019)]{kontar2019} Kontar, E.~P., Jeffrey, N.~L.~S., \& Emslie, A.~G.\ 2019, \apj, 871, 225

\bibitem[Krucker et al.(2014)]{krucker2014} Krucker, S., Christe, S., Glesener, L., et al.\ 2014, \apjl, 793, L32 

\bibitem[Kuhar et al.(2018)]{kuhar2018} Kuhar, M., Krucker, S., Glesener, L., et al.\ 2018, \apjl, 856, L32 

\bibitem[Neupert(1968)]{neupert1968} Neupert, W.~M.\ 1968, \apjl, 153, L59 

\bibitem[Ryan et al.(2014)]{ryan2014} Ryan, D.~F., O'Flannagain, A.~M., Aschwanden, M.~J., et al.\ 2014, \solphys, 289, 2547

\bibitem[Shih et al.(2019)]{shih2019} Shih, A.~Y., Glesener, L., Christe, S., et al.\ 2019, AGU Fall Meeting Abstracts 2019, SH33A-08

\bibitem[Veronig et al.(2002)]{veronig2002} Veronig, A., Vr{\v s}nak, B., Dennis, B.~R., et al.\ 2002, \aap, 392, 699 

\bibitem[Veronig \& Brown(2004)]{veronig2004} Veronig, A.~M., \& Brown, J.~C.\ 2004, \apjl, 603, L117 

\bibitem[Veronig et al.(2005)]{veronig2005} Veronig, A.~M., Brown, J.~C., \& Bone, L.\ 2005, Advances in Space Research, 35, 1683

\bibitem[Vievering et al.(2020)]{vievering2019} Vievering, J.T., et al., \apj, in prep.

\bibitem[Warmuth, \& Mann(2016)]{warmuth2016} Warmuth, A., \& Mann, G.\ 2016, \aap, 588, A116

\bibitem[Wright et al.(2017)]{wright2017} Wright, P.~J., Hannah, I.~G., Grefenstette, B.~W., et al.\ 2017, \apj, 844, 132 

\bibitem[Wright(2017)]{pjwright} Wright, P.~J. \ 2017, doi: {10.5281/zenodo.840393}, url: {https://github.com/PaulJWright/ColourBlind}

\end{thebibliography}
\end{document}